\PassOptionsToPackage{unicode}{hyperref}
\PassOptionsToPackage{hyphens}{url}
\PassOptionsToPackage{dvipsnames,svgnames,x11names}{xcolor}
\documentclass[
  11pt,
  11pt,
  letterpaper]{article}
\usepackage{xcolor}
\usepackage[margin=1in,top=1.25in,bottom=1.25in]{geometry}
\usepackage{amsmath,amssymb}
\usepackage{iftex}
\ifPDFTeX
  \usepackage[T1]{fontenc}
  \usepackage[utf8]{inputenc}
  \usepackage{textcomp} 
  \DeclareUnicodeCharacter{2212}{-} 
\else 
  \usepackage{unicode-math} 
  \defaultfontfeatures{Scale=MatchLowercase}
  \defaultfontfeatures[\rmfamily]{Ligatures=TeX,Scale=1}
\fi
\usepackage{lmodern}
\ifPDFTeX\else
\fi
\IfFileExists{upquote.sty}{\usepackage{upquote}}{}
\IfFileExists{microtype.sty}{
  \usepackage[]{microtype}
  \UseMicrotypeSet[protrusion]{basicmath} 
}{}
\usepackage{setspace}
\makeatletter
\@ifundefined{KOMAClassName}{
  \IfFileExists{parskip.sty}{%
    \usepackage{parskip}
  }{
    \setlength{\parindent}{0pt}
    \setlength{\parskip}{6pt plus 2pt minus 1pt}}
}{
  \KOMAoptions{parskip=half}}
\makeatother
\makeatletter
\ifx\paragraph\undefined\else
  \let\oldparagraph\paragraph
  \renewcommand{\paragraph}{
    \@ifstar
      \xxxParagraphStar
      \xxxParagraphNoStar
  }
  \newcommand{\xxxParagraphStar}[1]{\oldparagraph*{#1}\mbox{}}
  \newcommand{\xxxParagraphNoStar}[1]{\oldparagraph{#1}\mbox{}}
\fi
\ifx\subparagraph\undefined\else
  \let\oldsubparagraph\subparagraph
  \renewcommand{\subparagraph}{
    \@ifstar
      \xxxSubParagraphStar
      \xxxSubParagraphNoStar
  }
  \newcommand{\xxxSubParagraphStar}[1]{\oldsubparagraph*{#1}\mbox{}}
  \newcommand{\xxxSubParagraphNoStar}[1]{\oldsubparagraph{#1}\mbox{}}
\fi
\makeatother

\usepackage{longtable,booktabs,array}
\usepackage{calc} 
\usepackage{etoolbox}
\makeatletter
\patchcmd\longtable{\par}{\if@noskipsec\mbox{}\fi\par}{}{}
\makeatother
\IfFileExists{footnotehyper.sty}{\usepackage{footnotehyper}}{\usepackage{footnote}}
\makesavenoteenv{longtable}
\usepackage{graphicx}
\makeatletter
\newsavebox\pandoc@box
\newcommand*\pandocbounded[1]{
  \sbox\pandoc@box{#1}%
  \Gscale@div\@tempa{\textheight}{\dimexpr\ht\pandoc@box+\dp\pandoc@box\relax}%
  \Gscale@div\@tempb{\linewidth}{\wd\pandoc@box}%
  \ifdim\@tempb\p@<\@tempa\p@\let\@tempa\@tempb\fi
  \ifdim\@tempa\p@<\p@\scalebox{\@tempa}{\usebox\pandoc@box}%
  \else\usebox{\pandoc@box}%
  \fi%
}
\def\fps@figure{htbp}
\makeatother

\NewDocumentCommand\citeproctext{}{}

\makeatletter
 \let\@cite@ofmt\@firstofone
 \def\@biblabel#1{}
 \def\@cite#1#2{{#1\if@tempswa , #2\fi}}
\makeatother
\newlength{\cslhangindent}
\newlength{\csllabelwidth}
\newenvironment{CSLReferences}[2] 
 {\begin{list}{}{%
  \setlength{\itemindent}{0pt}
  \setlength{\leftmargin}{0pt}
  \setlength{\parsep}{0pt}
  \ifodd #1
   \setlength{\leftmargin}{\cslhangindent}
   \setlength{\itemindent}{-1\cslhangindent}
  \fi
  \setlength{\itemsep}{#2\baselineskip}}}
 {\end{list}}
\usepackage{calc}

\usepackage{microtype}
\usepackage{hyperref}
\hypersetup{
    colorlinks=true,
    linkcolor=blue,
    citecolor=blue,
    urlcolor=blue,
    pdfauthor={Mohammad Dastgheib and Fatemeh Pourmahdian},
    pdftitle={Adaptive Gaze-Hand Switching in XR: Modality-Specific Failure Modes and the Midas Touch Problem},
    pdfkeywords={Extended Reality, Virtual Reality, Adaptive Interfaces, Gaze Interaction, Fitts's Law, Human-Computer Interaction, Multimodal Interaction, Pupillometry}
}

\usepackage{amsmath}
\usepackage{amssymb}
\usepackage{graphicx}
\usepackage{booktabs}
\usepackage{tabularx}
\usepackage{orcidlink}
\usepackage{float}

\AtBeginEnvironment{table}{\small}

\usepackage{needspace}
\usepackage{etoolbox}
\BeforeBeginEnvironment{table}{\Needspace{10\baselineskip}}

\makeatletter
\renewcommand{\@maketitle}{%
  \newpage
  \null
  \vskip 2em%
  \begin{center}%
    \let \footnote \thanks
    {\LARGE \@title \par}%
    \vskip 1.5em%
    {\large
      \begin{tabular}[t]{c}%
        Mohammad Dastgheib\textsuperscript{1} \orcidlink{0000-0001-7684-3731} \\
        University of California, Riverside \\
        \href{mailto:mohammad.dastgheib@email.ucr.edu}{\texttt{mohammad.dastgheib@email.ucr.edu}} \\[0.5em]
        Fatemeh Pourmahdian \orcidlink{0009-0007-5473-2070} \\
        \href{mailto:fatemepourmahd@gmail.com}{\texttt{fatemepourmahd@gmail.com}} \\
        Independent Researcher \\
      \end{tabular}\par}%
  \end{center}%
  \vskip 0.5em%
  \begin{center}
    \footnotesize
    \begin{tabular}{@{}l@{}}
      \textsuperscript{1}Corresponding author: \href{mailto:mohammad.dastgheib@email.ucr.edu}{\texttt{mohammad.dastgheib@email.ucr.edu}}
    \end{tabular}
  \end{center}%
  \par
  \vskip 1.5em%
}
\makeatother
\makeatletter
\@ifpackageloaded{caption}{}{\usepackage{caption}}
\AtBeginDocument{%
\ifdefined\contentsname
  \renewcommand*\contentsname{Table of contents}
\else
  \newcommand\contentsname{Table of contents}
\fi
\ifdefined\listfigurename
  \renewcommand*\listfigurename{List of Figures}
\else
  \newcommand\listfigurename{List of Figures}
\fi
\ifdefined\listtablename
  \renewcommand*\listtablename{List of Tables}
\else
  \newcommand\listtablename{List of Tables}
\fi
\ifdefined\figurename
  \renewcommand*\figurename{Figure}
\else
  \newcommand\figurename{Figure}
\fi
\ifdefined\tablename
  \renewcommand*\tablename{Table}
\else
  \newcommand\tablename{Table}
\fi
}
\@ifpackageloaded{float}{}{\usepackage{float}}
\floatstyle{ruled}
\@ifundefined{c@chapter}{\newfloat{codelisting}{h}{lop}}{\newfloat{codelisting}{h}{lop}[chapter]}
\floatname{codelisting}{Listing}

\makeatother
\makeatletter
\@ifpackageloaded{caption}{}{\usepackage{caption}}
\@ifpackageloaded{subcaption}{}{\usepackage{subcaption}}
\makeatother
\usepackage{bookmark}
\IfFileExists{xurl.sty}{\usepackage{xurl}}{} 
\hypersetup{
  pdftitle={The Midas Touch in Gaze vs.~Hand Pointing: Modality-Specific Failure Modes and Implications for XR Interfaces},
  pdfauthor={Mohammad Dastgheib; Fatemeh Pourmahdian},
  colorlinks=true,
  linkcolor={blue},
  filecolor={Maroon},
  citecolor={blue},
  urlcolor={blue},
  pdfcreator={LaTeX via pandoc}}

\title{The Midas Touch in Gaze vs.~Hand Pointing: Modality-Specific
Failure Modes and Implications for XR Interfaces}
\author{Mohammad Dastgheib \and Fatemeh Pourmahdian}
\date{2026-03-09}
\begin{document}
\maketitle
\begin{abstract}
Extended Reality (XR) interfaces impose both ergonomic and cognitive
demands, yet current systems often force a binary choice between
hand-based input, which can produce fatigue, and gaze-based input, which
is vulnerable to the Midas Touch problem and precision limitations. We
introduce the xr-adaptive-modality-2025 platform, a web-based
open-source framework for studying whether modality-specific adaptive
interventions can improve XR-relevant pointing performance and reduce
workload relative to static unimodal interaction. The platform combines
physiologically informed gaze simulation, an ISO 9241-9 multidirectional
tapping task, and two modality-specific adaptive interventions: gaze
declutter and hand target-width inflation. We evaluated the system in a
2 × 2 × 2 within-subjects design manipulating Modality (Hand vs.~Gaze),
UI Mode (Static vs.~Adaptive), and Pressure (Yes vs.~No). Results from
N=69 participants show that hand yielded higher throughput than gaze
(5.17 vs.~4.73 bits/s), lower error (1.8\% vs.~19.1\%), and lower
NASA-TLX workload. Crucially, error profiles differed sharply by
modality: gaze errors were predominantly slips (99.2\%), whereas hand
errors were predominantly misses (95.7\%), consistent with the Midas
Touch account. Of the two adaptive interventions, only gaze declutter
executed in this dataset; it modestly reduced timeouts but not slips.
Hand width inflation was not evaluable due to a UI integration bug.
These findings reveal modality-specific failure modes with direct
implications for adaptive policy design, and establish the platform as a
reproducible infrastructure for future studies.
\end{abstract}

\setstretch{1}
\vspace{-0.5em}


\newpage

\section{Introduction}\label{introduction}

XR interaction changes the control demands of pointing by moving input
from desktop devices to embodied hand and gaze actions. Unlike the
desktop metaphor, where interaction is mediated by low-effort devices
like the mouse and keyboard, XR requires the user to engage their entire
body; the primary pointing devices are the user's own hands and eyes.
This embodied interaction is fraught with ergonomic and cognitive
challenges. The ``Gorilla Arm'' syndrome, a phenomenon where prolonged
mid-air arm extension leads to rapid musculoskeletal fatigue and pain,
remains a critical barrier to the long-term adoption of gestural
interfaces. Conversely, gaze-based interaction, which leverages the
speed of the human oculomotor system, suffers from the ``Midas Touch''
problem---the inherent ambiguity between looking for perception and
looking for action---and lacks the fine motor precision required for
granular manipulation tasks (Jacob, 1990).

Current XR systems typically force a binary choice: the user must either
commit to a controller-based paradigm, accepting the physical fatigue,
or a gaze-based paradigm, accepting the lack of precision and the
potential for inadvertent triggers. This rigid dichotomy ignores the
dynamic nature of human attention and the varying demands of different
tasks. A high-precision manipulation task may require the stability of a
hand controller, while a rapid visual search task is best served by the
saccadic speed of the eye.

To investigate whether adaptive support can alleviate these tradeoffs,
we introduce the xr-adaptive-modality-2025 platform, a research
framework for studying modality-specific adaptive support within hand
and gaze input in XR-relevant pointing tasks. The system combines an ISO
9241-9 pointing task, physiologically informed gaze simulation, and
modality-specific adaptive interventions intended to address distraction
during gaze selection and spatial difficulty during hand selection. The
central question is whether modality-specific adaptive support can
improve performance and workload relative to static unimodal interaction
(Soukoreff \& MacKenzie, 2004).

This paper makes three contributions. First, we present an
\textbf{open-source platform} for controlled, reproducible study of gaze
and hand pointing in XR-relevant tasks, including a physiologically
informed gaze simulation, a policy-driven adaptation engine, and a full
remote data collection infrastructure. Second, we provide
\textbf{quantitative empirical evidence} on gaze vs.~hand input in an
ISO 9241-9 pointing task (N=69), showing that hand substantially
outperforms gaze on throughput and error rate, and that the dominant
failure modes are modality-specific: slips (false activations) for gaze
and misses (spatial targeting failures) for hand. Third, we demonstrate
that a simple \textbf{gaze-declutter adaptation} modestly reduces
timeouts but does not address the core slip-based failure mode of gaze
interaction, which has direct implications for designing
modality-specific adaptive interventions. Unlike prior gaze+hand
paradigms that treat the two modalities as complementary components of a
fixed multimodal combination (e.g., gaze for acquisition + hand for
confirmation (Jacob, 1990)), this work investigates
\emph{modality-specific adaptive support}---when and how the system
should activate adaptations (declutter, width inflation) within each
modality based on real-time performance signals. The platform could
support future studies of modality switching.

The study is guided by three research questions. \textbf{RQ1
(Performance):} Does a context-aware adaptive system yield higher
throughput (\(TP\)) than static unimodal systems? \textbf{RQ2
(Workload):} Can modality-specific adaptive interventions reduce
``Physical Demand'' and ``Frustration'' (NASA-TLX) compared to static
conditions? \textbf{RQ3 (Adaptation):} Do adaptive interventions improve
performance and reduce workload relative to static conditions? (In this
dataset, only gaze declutter was evaluable; hand width inflation was not
applied due to a UI integration bug.)

\section{Background and Related Work}\label{background-and-related-work}

\subsection{Sensorimotor Implications of Hand and Gaze
Input}\label{sensorimotor-implications-of-hand-and-gaze-input}

To design an effective adaptive system, one must first deconstruct the
physiological mechanisms of the component modalities. Hand and gaze
afford different control properties in XR: hand input supports precise
corrective control, whereas gaze supports rapid orienting but introduces
ambiguity when fixation is used for selection.

\textbf{Manual Pointing in XR:} Manual input in XR, whether through held
controllers or optical hand tracking, mimics the act of physical
pointing. This interaction style benefits from proprioception---the
body's innate sense of limb position---which allows for high-precision
corrections without visual attention. However, the biomechanical cost is
substantial. In a 1:1 mapped XR environment, reaching a virtual object
requires a corresponding physical motion. Frequent large-amplitude
movements lead to fatigue in the deltoids and trapezius muscles. As
fatigue sets in, the signal-to-noise ratio of the motor system degrades;
the hand begins to tremor, increasing the effective target width
required for accurate selection and reducing the overall throughput of
the interaction.

\textbf{Gaze Interaction:} The eye is the fastest motor organ in the
human body. Saccades---rapid, ballistic movements of the eye---can reach
velocities exceeding 900 degrees per second (Bahill \& Stark, 1979),
making gaze an incredibly efficient modality for target acquisition.
However, the eye is fundamentally an input organ, not an output device.
Using gaze for selection introduces several critical issues: (1)
\textbf{The Midas Touch}---in the physical world, we can look at an
object without interacting with it, but in a gaze-controlled interface,
looking becomes equivalent to touching, requiring ``dwell time''
mechanisms that slow interaction (Jacob, 1990); (2)
\textbf{Microsaccades and Jitter}---even when ``fixated,'' the eye
performs microsaccades to refresh the retinal image (Martinez-Conde et
al., 2004), meaning a gaze cursor is inherently noisy, making selection
of small targets frustrating without smoothing algorithms; (3)
\textbf{Saccadic Suppression}---during rapid eye movements, the visual
system suppresses input to prevent motion blur (Bridgeman et al., 1975),
creating a ``blind'' phase that makes the initial phase of gaze
targeting effectively open-loop.

\subsection{Signal Processing and Cognitive Load
Framing}\label{signal-processing-and-cognitive-load-framing}

To study these dynamics with experimental rigor and reproducibility, our
platform employs a \textbf{psychophysically-grounded generative
simulation} of gaze behavior rather than raw sensor input or hardware
eye-tracking. Unlike raw eye-tracker signals---which conflate
device-specific artifacts with genuine oculomotor dynamics---this
simulation provides explicit parametric control over the three
physiological constraints most relevant to dwell-based selection: (1)
\textbf{Sensor Lag}---a first-order lag (linear interpolation) mimics
the processing latency (30--70 ms) typical of video-based eye trackers
(Saunders \& Woods, 2014); (2) \textbf{Saccadic Blindness}---the cursor
is frozen during high-velocity movements (\textgreater120 deg/s),
simulating the lack of visual feedback during a saccade (Bridgeman et
al., 1975); (3) \textbf{Fixation Jitter}---Gaussian noise
(\(\sigma \approx\) 0.12°) is injected at low velocities to mimic
fixational drift and microsaccade statistics (Martinez-Conde et al.,
2004; Rolfs, 2009), ensuring that the cost of gaze interaction is
accurately and reproducibly represented. Because each parameter is
grounded in oculomotor literature, the simulation functions as a
controlled analog of real eye-tracking rather than an arbitrary proxy
(see Figure~\ref{fig-psychophysics} in Methods).

We use Cognitive Load Theory (CLT) as a conceptual lens for reasoning
about modality-specific costs (Sweller, 1988). CLT distinguishes
intrinsic load (task difficulty) from extraneous load (interface
overhead). In principle, hand input may increase physical effort
(intrinsic load during larger reaches), whereas gaze input may increase
extraneous attentional and verification demands during precise
selection. The present study uses workload (NASA-TLX) and performance
measures to test whether these modality-specific costs appear in the
current task, and whether adaptive interventions can reduce them.

\subsection{Adaptive Intervention
Mechanisms}\label{adaptive-intervention-mechanisms}

We implemented two modality-specific adaptive interventions. \textbf{For
gaze interaction}, a declutter mechanism draws on visual attention
guiding techniques in XR, related to diminished reality (DR) (Herling \&
Broll, 2010). When the policy engine detects performance degradation in
gaze mode (error burst or RT threshold exceeded), non-critical HUD
elements are hidden, mitigating peripheral distraction. This aligns with
foveated rendering principles, where systems leverage the human visual
system's foveal focus to prioritize content (Patney et al., 2016). By
decluttering when performance degrades, we hypothesize that gaze-based
targeting becomes faster and less cognitively demanding.

\textbf{For hand-based interactions}, our adaptive strategy is width
inflation---dynamically expanding the effective size of targets when
performance degrades, triggered by the policy engine based on error rate
and reaction time thresholds. This concept is inspired by ``expanding
targets'' research (McGuffin \& Balakrishnan, 2005) and the ``Bubble
Cursor'' (Grossman \& Balakrishnan, 2005), which demonstrate that even
slight enlargement significantly improves pointing performance. Our
implementation uses a hysteresis gate (requiring N consecutive trials
meeting trigger conditions) to prevent flicker, acting as a ``safety
net'' that compensates for motor tremor under fatigue.

\subsection{Differentiation from Prior Gaze+Hand
Work}\label{differentiation-from-prior-gazehand-work}

Recent XR systems combine gaze and hand in fixed multimodal
combinations: gaze for target acquisition and hand for confirmation or
manipulation (Pfeuffer et al., 2017, 2024). Pfeuffer et al. (2024)
outline design principles for gaze+pinch interaction (e.g., division of
labor, minimalistic timing), while Kim et al. (2025) propose
PinchCatcher, a semi-pinch quasi-mode for multi-selection in gaze+pinch
interfaces. These approaches treat gaze and hand as complementary
components of a fixed interaction paradigm. Our work differs by
investigating \emph{modality-specific adaptive support}---when and how
the system should activate adaptations within each modality based on
real-time performance signals---rather than a fixed multimodal
combination. We draw on context-aware computing principles (Dey, 2001)
to frame policy-driven adaptation that responds to performance
degradation.

\section{Methods}\label{methods}

The study was designed to support reproducible evaluation of
modality-specific adaptive interventions under controlled remote testing
conditions. The platform \texttt{xr-adaptive-modality-2025} serves as
the technical apparatus for the study.

\subsection{Apparatus}\label{apparatus}

We developed a custom pointing testbed as a web-based application (React
18, TypeScript), allowing broad hardware compatibility for remote
participants. The study was conducted on participants' own computers
using a standard mouse or trackpad. No headset or VR hardware was used;
the platform serves as a controlled web-based proxy for XR-relevant
selection dynamics.

\subsubsection{Display Calibration and Reliability
Measures}\label{display-calibration-and-reliability-measures}

To ensure measurement validity across heterogeneous display
configurations, we implemented a multi-layered approach addressing
display variability. Before commencing experimental trials, participants
completed a Credit Card Calibration procedure: participants placed a
standard credit card (85.60 mm × 53.98 mm) on their screen and adjusted
an on-screen rectangle to match its physical dimensions. This
calibration enabled computation of pixels per millimeter (px/mm) and
pixels per degree of visual angle (PPD), normalizing gaze simulation
jitter to screen-space pixels and ensuring consistent perceptual
difficulty across different display sizes and resolutions (MacKenzie,
1992).

To minimize measurement error, we enforced strict display requirements:
fullscreen/maximized window (required before starting blocks), browser
zoom locked to 100\% (verified before each block using
\texttt{window.visualViewport.scale}), and live monitoring during trials
(trials automatically paused if settings changed). For every trial, we
logged comprehensive display metadata: device pixel ratio (DPR), browser
type, viewport dimensions, zoom level, fullscreen status, and tab
visibility duration. Trials were excluded from analysis if zoom level
\(\neq\) 100\%, fullscreen status = FALSE, DPR instability (change
\textgreater{} 0.1 between blocks), or tab hidden for \textgreater{}
500ms. Participants with \textgreater{} 40\% of trials excluded due to
display violations were removed from the final analysis.

\subsubsection{Gaze Simulation}\label{gaze-simulation}

To ensure rigorous internal validity and precise control over noise
characteristics, we utilized a Physiologically-Accurate Gaze Simulation.
The simulation models the specific constraints of eye-tracking
interaction---latency and jitter---with explicit parametric control. The
simulation transformed raw mouse input into ``gaze'' coordinates via
three mechanisms derived from oculomotor physiology:

\begin{enumerate}
\def\labelenumi{\arabic{enumi}.}
\item
  \textbf{Saccadic Suppression \& Ballistic Movement:} The cursor was
  ``frozen'' (blind) during high-velocity movements (\textgreater120
  deg/s) to simulate the brain's suppression of visual input during
  saccades, a phenomenon known as saccadic suppression of image
  displacement (Bridgeman et al., 1975). This aligns with the ballistic
  nature of saccadic eye movements, where visual feedback is effectively
  open-loop until the eye settles.
\item
  \textbf{Fixation Jitter \& Drift:} When the cursor slowed (\textless30
  deg/s), Gaussian noise (SD \(\approx\) 0.12\(^\circ\) visual angle)
  was injected to simulate fixational eye movements (Martinez-Conde et
  al., 2004), specifically the random walk characteristics of ocular
  drift and tremor that occur even during attempted fixation. This
  angular noise was normalized to screen pixels using the
  pixels-per-degree (PPD) calibration value, ensuring consistent
  perceptual difficulty across different display sizes and viewing
  distances.
\item
  \textbf{Sensor Lag:} A first-order lag (linear interpolation, factor
  0.15) was applied to mimic the processing latency (typically 30--70
  ms) inherent in video-based eye trackers (Saunders \& Woods, 2014).
\end{enumerate}

\begin{figure}[H]

\centering{

\includegraphics[width=1\linewidth,height=\textheight,keepaspectratio]{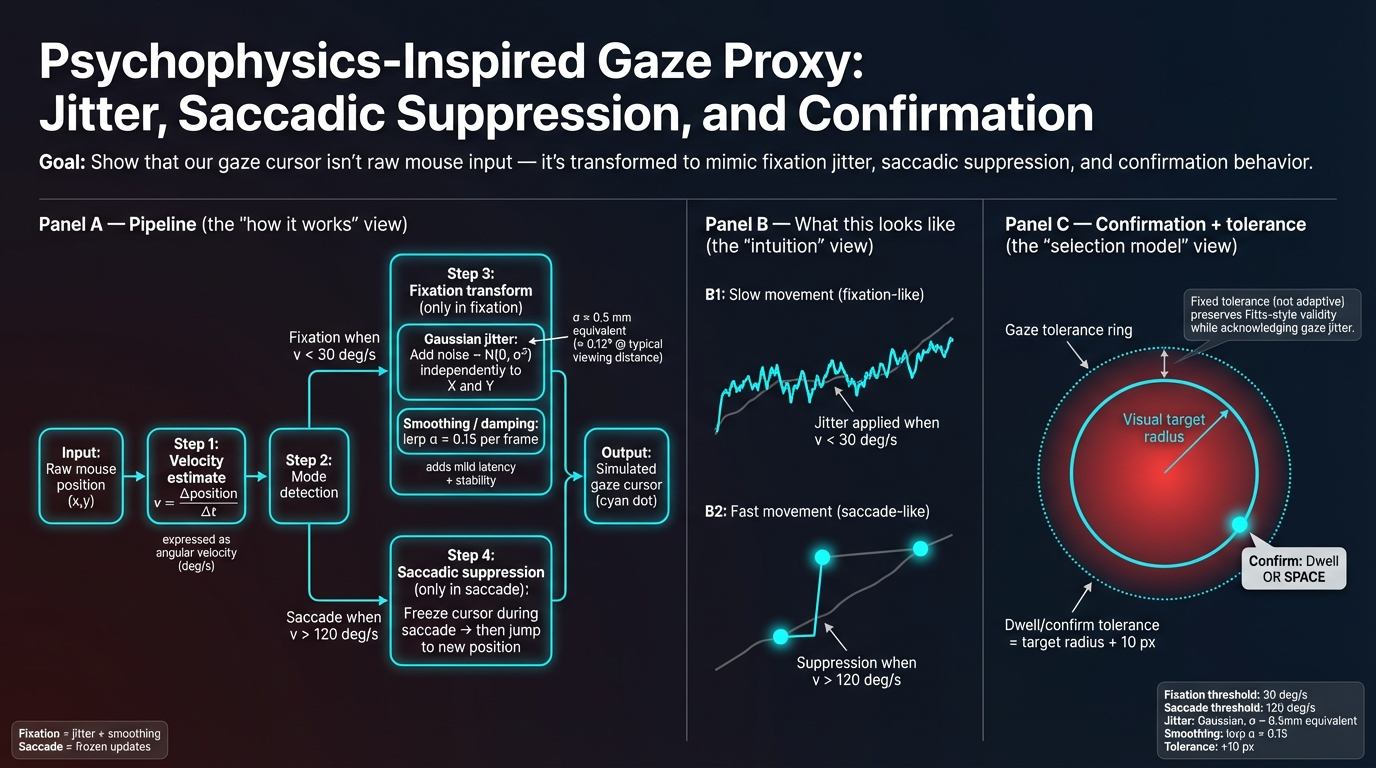}

}

\caption{\label{fig-psychophysics}Psychophysically-grounded gaze proxy
pipeline. Panel A (left): The three-stage simulation: raw mouse input is
differentiated to estimate angular velocity; velocity above 120 deg/s
triggers saccadic suppression (cursor freezes, then jumps to new
position); velocity below 30 deg/s triggers the fixation transform
(Gaussian jitter \(\sigma \approx\) 0.5 mm + first-order lag smoothing).
Panel B (center): Output signal examples---fixation mode (B1) produces
continuous spatial noise consistent with microsaccade statistics;
saccade mode (B2) produces a cursor freeze and ballistic jump,
reproducing the perceptual blind phase of real saccades. Panel C
(right): The selection model, with a dwell-confirm tolerance ring
(target radius + 10 px) that accommodates fixation jitter while keeping
sensitivity well-defined.}

\end{figure}

Each parameter was chosen to match published oculomotor norms, making
the simulation a psychophysically-grounded proxy for eye-tracking rather
than arbitrary noise injection. The 120 deg/s saccade-detection
threshold lies well above the velocity range of fixation and smooth
pursuit (\textless30 deg/s) and below the peak velocities of
medium-to-large saccades (\textgreater300 deg/s), consistent with
established velocity-based saccade classification criteria (Bahill \&
Stark, 1979; Engbert \& Kliegl, 2003). During saccades in this velocity
range, the visual system suppresses sensitivity by up to 50--80\% of
baseline, preventing perceptual blur from rapid retinal image
displacement (Bridgeman et al., 1975). The fixation jitter magnitude
(\(\sigma \approx\) 0.12°) falls within the documented amplitude range
of human microsaccades---the involuntary fixational eye movements
responsible for the dominant source of spatial noise at a gaze
cursor---which span 0.1--0.5° in the literature (Martinez-Conde et al.,
2004; Rolfs, 2009). Applied independently per axis per frame, this
produces a spatial random walk consistent with the statistical
characteristics of fixational drift and tremor. The sensor lag (lerp
\(\alpha\) = 0.15 at 60 Hz) produces an effective smoothing time
constant of approximately 100 ms, encompassing the hardware plus
software latency range documented for commercial video-based eye
trackers (Saunders \& Woods, 2014). Figure~\ref{fig-psychophysics}
illustrates the full pipeline, the behavioral output of each stage, and
the selection tolerance model.

\subsection{Input Modality
Implementations}\label{input-modality-implementations}

The system implemented two distinct input modalities, each with static
and adaptive modes. \textbf{For hand-based trials}, participants
controlled a cursor with their mouse to click the target. In static
mode, this was standard 1:1 mouse pointing. In adaptive mode, the system
was designed to expand the effective clickable area of targets (width
scale \textgreater{} 1.0) through a rule-based policy engine that
triggers when performance degrades (error burst \(\geq\) 2 consecutive
errors or reaction time exceeds the 75th percentile threshold), drawing
on the ``expanding targets'' technique (McGuffin \& Balakrishnan, 2005).
In this dataset, the hand width-inflation pathway was not evaluable:
width scaling remained at 1.0 across trials, and although the policy
engine emitted inflate-width actions in some sessions, the rendered
targets were not updated.

\textbf{For gaze-based trials}, participants controlled the cursor via
the simulated gaze signal described above. In the analyzed dataset, gaze
selection used 500 ms dwell-to-select: the cursor had to remain within
the target for 500 ms to trigger selection automatically. Space key
confirmation (dwell disabled) was not used. This follows the standard
``dwell-to-select'' paradigm (Majaranta et al., 2006; Ware \& Mikaelian,
1987). In adaptive mode, the system implements a \emph{declutter
mechanism} that hides non-critical HUD elements when the policy engine
detects performance degradation (error burst or RT threshold exceeded)
in gaze blocks. The declutter effect persists until performance
improves, using hysteresis to prevent rapid oscillation.

Both adaptive features (declutter and width expansion) are driven by a
rule-based policy engine with hysteresis gates (requiring N consecutive
trials meeting trigger conditions) to prevent flicker. The application
logged detailed event traces (policy state changes, trial performance,
adaptation triggers) to enable verification of adaptive mechanism
activation and effectiveness.

\subsection{Task and Stimuli}\label{task-and-stimuli}

\begin{figure}[H]

\centering{

\includegraphics[width=1\linewidth,height=\textheight,keepaspectratio]{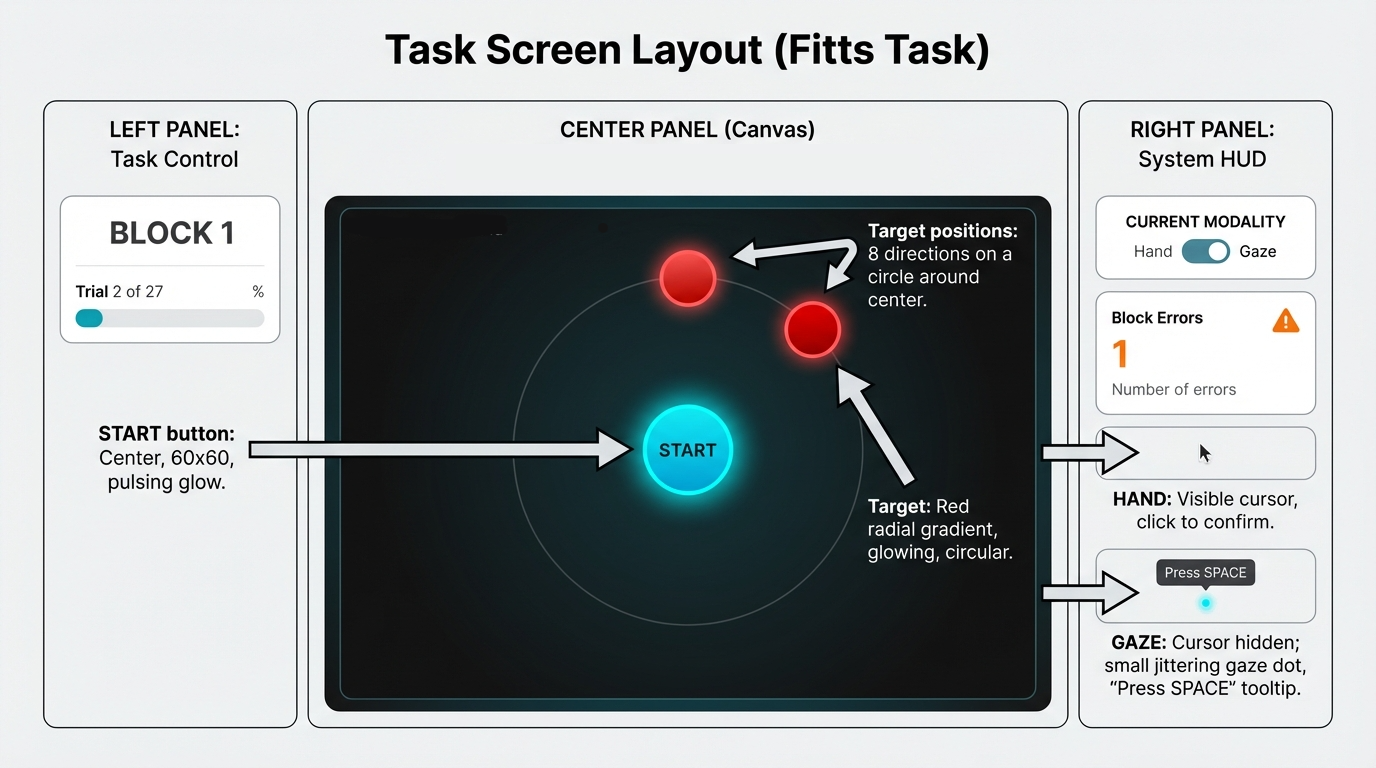}

}

\caption{\label{fig-task-layout}Task UI overview. ISO 9241-9
multi-directional tapping task: participants select highlighted targets
on a central canvas. A side HUD shows modality and block-level
feedback.}

\end{figure}

Participants performed a multi-directional pointing task conforming to
the ISO 9241-9 standard for non-keyboard input device evaluation (ISO
9241-9, 2000). Targets were arranged in a circular layout with 8
positions (width \(W\), amplitude \(A\)), with one target highlighted at
a time. Targets were presented with IDs ranging from approximately 2 to
6 bits, calculated using the Shannon formulation of Fitts' Law (target
width ranged from 30 px to 80 px, with corresponding distances chosen to
yield the desired ID values). In half of the blocks, a ``Time Pressure''
condition was enforced via a visible countdown timer; failure to select
within the timeout (6s) resulted in a forced error, intended to induce
stress and mental workload, simulating a high-demand scenario.
Figure~\ref{fig-task-layout} shows the task interface.

\subsection{Experimental Design}\label{experimental-design}

We employed a repeated-measures factorial design: all participants
experienced every combination of the two input modalities (Gaze
vs.~Hand) × two UI conditions (Adaptive vs.~Non-adaptive) × two workload
levels (Pressure vs.~No Pressure). This creates a 2 × 2 × 2
within-subjects design.

\subsubsection{Counterbalancing: The Williams
Design}\label{counterbalancing-the-williams-design}

The order of modality blocks was counterbalanced using a Williams Latin
square arrangement to control for learning effects (Williams, 1949).
Static and adaptive conditions were run as \textbf{separate blocks};
each block corresponded to one Modality × UI Mode × Pressure combination
(e.g., Hand-Static-Self-Paced, Gaze-Adaptive-Time-Limited). Within each
block, all trials shared the same modality, UI mode, and pressure level.
Participants were not explicitly told when the system was adapting,
aside from noticing the visual changes, to reduce expectancy biases.

\subsection{Participants}\label{participants}

\subsubsection{Sample Size and
Recruitment}\label{sample-size-and-recruitment}

A total of 81 participants enrolled and completed the task. After
applying the inclusion criteria, study's trial-validity, and
factorial-completeness criteria, 69 participants were retained for the
primary analysis. The retained sample had a mean age of 30.0 years (SD =
7.5; range = 18--62) and included 39 men and 30 women. Most participants
were right-handed (91.3\%), and all retained participants completed the
hand condition using a mouse with the right hand. Participants reported
no motor impairments. This study was conducted as an independent
personal research project and was not reviewed by an institutional
review board. Participants provided informed consent electronically
through the web application before participation.

\subsection{Procedure}\label{procedure}

Each participant completed a short training session to get familiar with
gaze selection (including practice with the simulated gaze interface and
dwell clicking) and hand selection. During the experiment, they
performed 8 blocks of 27 trials each (216 main-task trials per
participant before trial-level exclusions). Practice trials were
excluded from analysis. Target positions cycled through 8 directions;
Index of Difficulty varied across three levels (\(\approx\) 2--6 bits).
Pressure (time-limited vs.~self-paced) was assigned at the block level.

After each block, participants filled out a NASA-TLX workload survey
(rating mental demand, physical demand, etc.) and took a short break to
mitigate fatigue. The entire session lasted about 1 hour per
participant.

\subsection{Measures and Analysis
Strategy}\label{measures-and-analysis-strategy}

The primary performance measures were Movement Time (MT) in milliseconds
(from trial start to successful selection) and Selection Accuracy (hit
vs.~miss rate, including specific error types). We classified errors
into three categories following Reason's (1990) taxonomy of human error:
\textbf{slips} (unintended activations---the cursor dwelled on a target
the user did not intend to select), \textbf{misses} (failures of spatial
targeting---the user attempted selection but did not acquire the target
within bounds), and \textbf{timeouts} (no response within the 6-second
trial window). The main text emphasizes descriptive estimates and 95\%
confidence intervals.

\subsubsection{Throughput and Fitts-Style Performance
Metrics}\label{throughput-and-fitts-style-performance-metrics}

We adopted the Shannon formulation of Fitts's Law (Soukoreff \&
MacKenzie, 2004), the standard form used for ISO 9241-9 compliance:
\(ID = \log_2(D/W + 1)\), where \(D\) is target distance and \(W\) is
target width. Movement time was modeled as \(MT = a + b \cdot ID\),
where \(a\) is the intercept and \(b\) reflects the movement-time cost
per additional bit of difficulty; smaller slopes therefore imply greater
information-processing efficiency. To account for endpoint variability,
we used Effective Width (\(W_e\)), computed from the distribution of
selection endpoints as \(W_e = 4.133 \sigma_x\), where \(\sigma_x\) is
the standard deviation of endpoint error projected onto the task axis.
This ISO-based correction yields an effective index of difficulty
(\(ID_e\)) and allows Throughput (\(TP = ID_e / MT\)) to serve as a
unified measure of speed--accuracy performance. We also logged velocity
profiles and submovement counts for control-theoretic analysis (Meyer et
al., 1988), although those measures are not the focus of the present
report.

\subsubsection{Verification-Phase Modeling with
LBA}\label{verification-phase-modeling-with-lba}

To better understand behavior after the pointer first entered the
target, we modeled the verification phase---the interval between first
target entry and final selection---as a latent decision process. This
phase is intended to capture the extra time users may spend confirming
that the correct target has been acquired before committing to
selection. We used the \textbf{Linear Ballistic Accumulator (LBA)} model
(Brown \& Heathcote, 2008), a race model in which evidence for competing
responses increases linearly until one reaches threshold. In behavioral
terms: \textbf{drift rate} reflects how quickly and how well evidence
accumulates that the target acquisition is correct; \textbf{threshold}
reflects how much evidence is required before committing to selection;
and \textbf{\(t_0\)} is a verification-related latent offset capturing
residual non-accumulation time.

In our fitted parameterization, \(t_0\) is reported on a latent scale
rather than as a directly observed millisecond quantity, so condition
differences in \(t_0\) are interpreted directionally and in conjunction
with the empirical verification-phase RT summaries. We chose LBA rather
than a Drift Diffusion Model because these data come from a low-error
pointing task, a setting in which LBA is often more stable and
interpretable than diffusion-based approaches (Lerche et al., 2017). We
fit a hierarchical Bayesian LBA model in PyMC with modality- and
UI-mode-varying \(t_0\), ID-varying drift rate, and pressure-varying
threshold, using verification-phase RTs from valid trials (200--5000
ms).

\subsection{Deployment and Data Collection
Infrastructure}\label{deployment-and-data-collection-infrastructure}

The experimental platform was deployed as a web application (React 18,
TypeScript, Vite) hosted on Vercel
(https://xr-adaptive-modality-2025.vercel.app) to enable remote,
asynchronous data collection. Each participant received a unique URL
containing embedded participant ID. The application automatically
detected these parameters on load, initializing the session with the
appropriate identifier. Session state was managed client-side using
browser localStorage, enabling participants to pause and resume sessions
while maintaining progress tracking.

Data collection occurred entirely client-side to ensure participant
privacy. The application implemented a structured CSV logging system
that captured comprehensive trial-level data in real-time (77 columns
per trial), including participant metadata (ID, demographics, session
number), trial parameters (block, trial, modality, UI mode, pressure,
ID, A, W), performance metrics (RT, accuracy, error type, hover
duration, submovement count, verification time), adaptive system metrics
(width scaling, alignment gate metrics, adaptation triggers), system
metadata (browser type, DPR, display calibration, timestamp), and
workload measures (NASA-TLX subscales). At the completion of each
session, participants exported their data via browser-based CSV
download. Data export occurred entirely locally---no data was
transmitted to servers during the experimental session, ensuring
participant privacy. The application was built as a single-page
application (SPA) with client-side routing, with the gaze simulation
algorithm, adaptive policy engine, and data logging system all operating
in real-time within the browser. Event-driven architecture (via an
internal event bus) coordinated trial timing, data logging, and UI
updates, ensuring precise temporal alignment between user actions and
recorded data.

\textbf{Data Quality Assurance:} A post-collection audit verified
modality and UI mode logging and identified a pressure-condition logging
issue that affected early sessions. The bug was corrected in the
codebase; all analyses use the corrected merged dataset. The primary
participant exclusion criterion is 8-block factorial completeness. All
code is open-source and available for reproducibility.

\section{Results}\label{results}

We report results from N=69 participants with complete factorial data
(15,105 trials after QC and device filter; 13,519 valid for performance
metrics).

\subsection{Primary Performance Outcomes
(RQ1)}\label{primary-performance-outcomes-rq1}

Throughput (TP), error rate, and movement time (MT) were computed
following ISO 9241-9. Table~\ref{tbl-performance} reports descriptive
statistics by modality, collapsed over UI mode and pressure.

\begin{longtable}[]{@{}lcc@{}}
\caption{Primary performance metrics by modality. Values are mean
{[}95\% CI{]}. Hand produced higher throughput and lower error rate than
gaze.}\label{tbl-performance}\tabularnewline
\toprule\noalign{}
Metric & Hand & Gaze \\
\midrule\noalign{}
\endfirsthead
\toprule\noalign{}
Metric & Hand & Gaze \\
\midrule\noalign{}
\endhead
\bottomrule\noalign{}
\endlastfoot
Throughput (bits/s) & 5.17 {[}5.06, 5.27{]} & 4.73 {[}4.58, 4.88{]} \\
Error Rate (\%) & 1.77 {[}1.22, 2.32{]} & 19.09 {[}17.58, 20.59{]} \\
Movement Time (s) & 1.09 {[}1.07, 1.11{]} & 1.19 {[}1.14, 1.23{]} \\
\end{longtable}

\begin{figure}[H]

\centering{

\includegraphics[width=0.85\linewidth,height=\textheight,keepaspectratio]{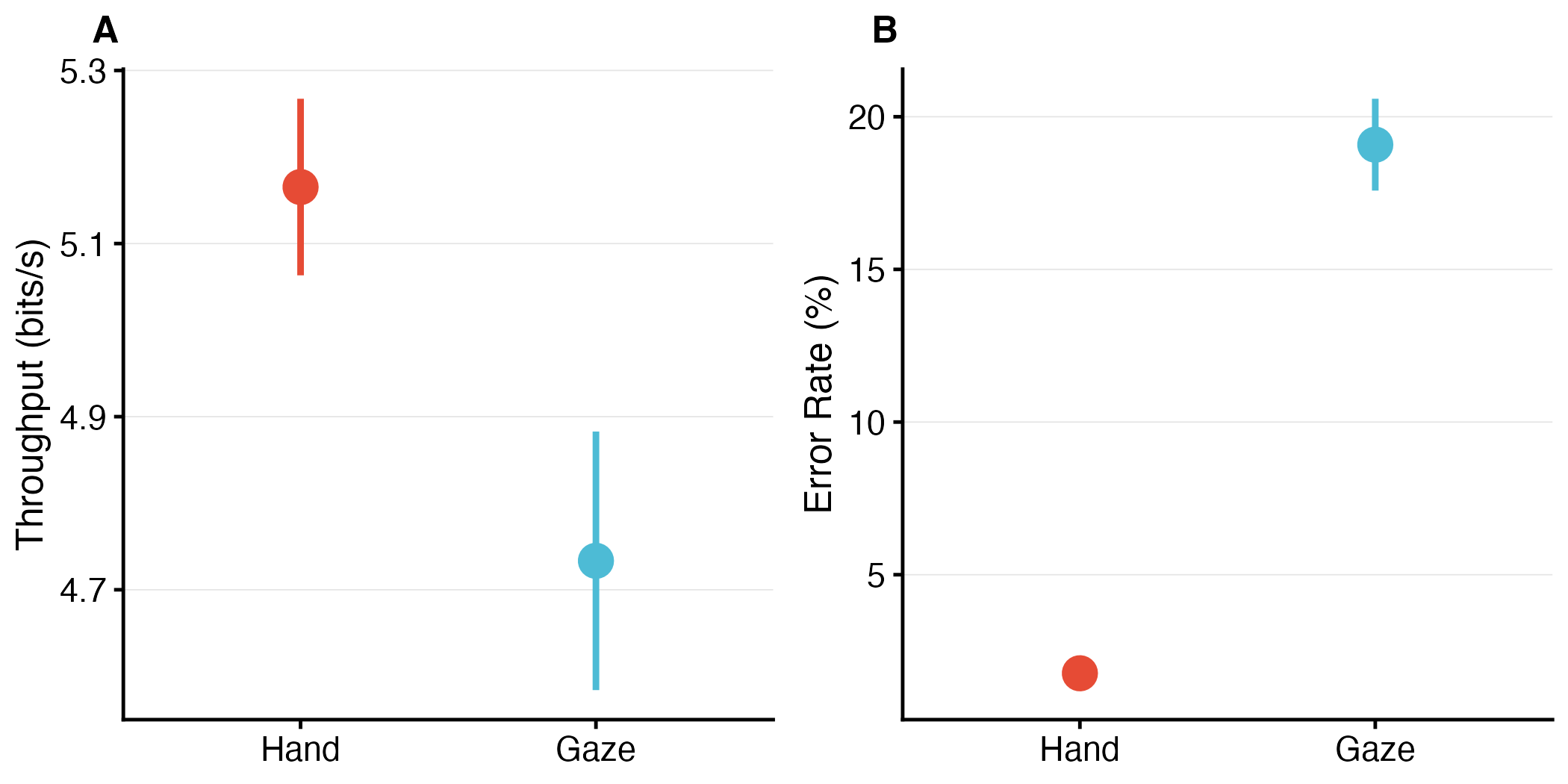}

}

\caption{\label{fig-performance}Primary performance by modality. (A)
Throughput (bits/s): hand produced higher throughput than gaze. (B)
Error rate (\%): gaze showed substantially higher error rate than hand.
Error bars show 95\% CI.}

\end{figure}

Hand input yielded higher throughput (5.17 vs.~4.73 bits/s) and
substantially lower error rate (1.8\% vs.~19.1\%) than gaze input.
Movement time was shorter for hand (1.09 s) than gaze (1.19 s). Hand
outperformed gaze on all three metrics. Figure~\ref{fig-performance}
illustrates these differences.

\subsection{Error Profile: Midas Touch
Signature}\label{error-profile-midas-touch-signature}

Error types differed sharply between modalities
(Table~\ref{tbl-error-types}). For \textbf{gaze}, 99.2\% of errors were
\textbf{slips} (accidental activations) and 0.8\% were timeouts. For
\textbf{hand}, 95.7\% were \textbf{misses} (target not acquired) and
4.3\% were timeouts. This asymmetry is consistent with the Midas Touch
account (Jacob, 1990): gaze interaction failed primarily due to intent
ambiguity (looking to see vs.~looking to select), whereas hand
interaction failed due to spatial targeting errors.

\begin{longtable}[]{@{}lcc@{}}
\caption{Distribution of error types by modality. Gaze errors are
predominantly slips; hand errors are predominantly
misses.}\label{tbl-error-types}\tabularnewline
\toprule\noalign{}
Error Type & Hand & Gaze \\
\midrule\noalign{}
\endfirsthead
\toprule\noalign{}
Error Type & Hand & Gaze \\
\midrule\noalign{}
\endhead
\bottomrule\noalign{}
\endlastfoot
Slip (accidental activation) & 0\% & 99.2\% \\
Miss (target not acquired) & 95.7\% & 0\% \\
Timeout & 4.3\% & 0.8\% \\
\end{longtable}

\begin{figure}[H]

\centering{

\includegraphics[width=0.75\linewidth,height=\textheight,keepaspectratio]{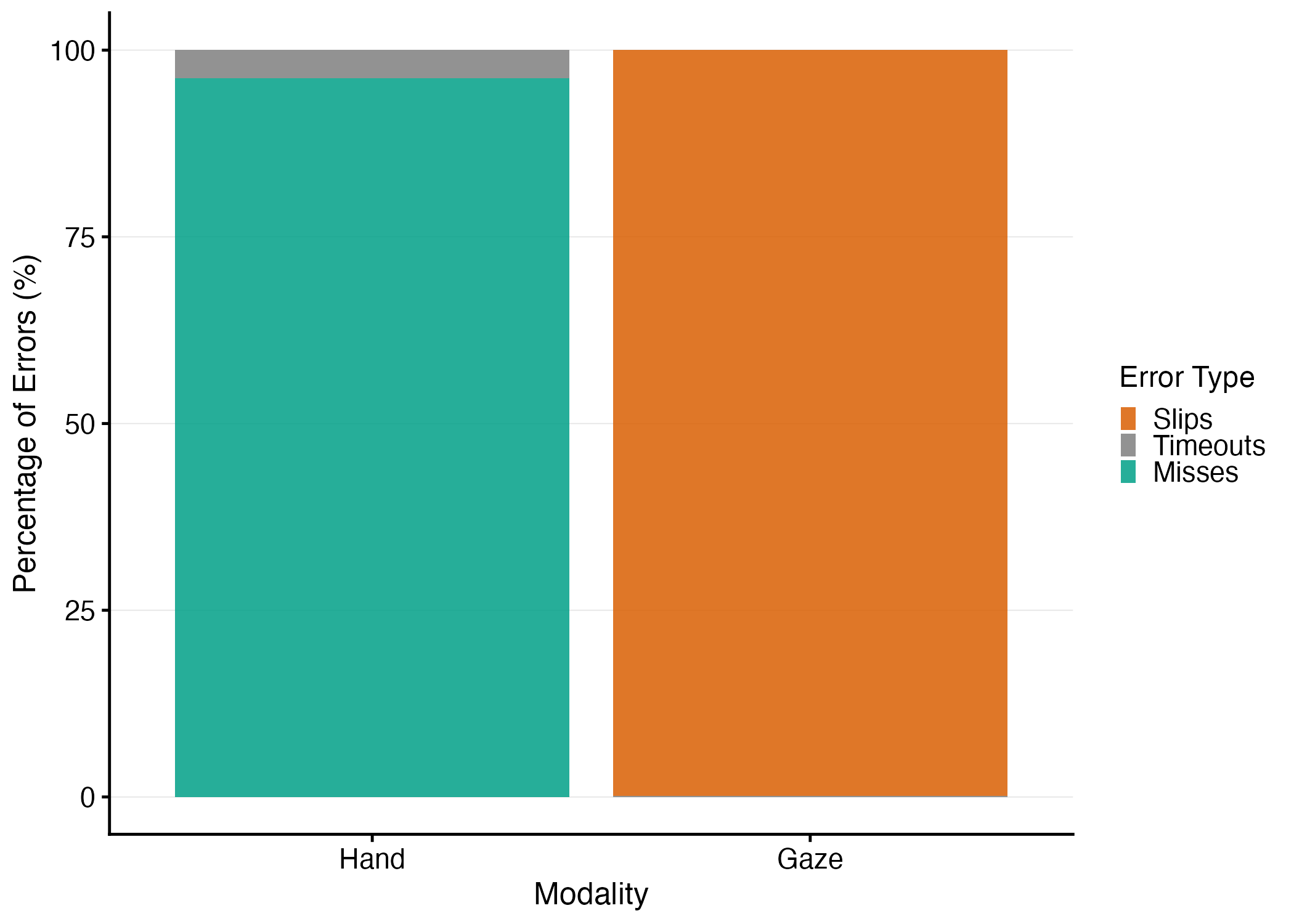}

}

\caption{\label{fig-error-types}Error type composition by modality. Gaze
errors are predominantly slips (accidental activations); hand errors are
predominantly misses.}

\end{figure}

Figure~\ref{fig-error-types} shows the composition of errors by
modality, illustrating the stark contrast between gaze (slips) and hand
(misses).

\subsection{Gaze Declutter Effectiveness
(RQ3)}\label{gaze-declutter-effectiveness-rq3}

The gaze adaptive manipulation (declutter) was the only adaptation that
executed in this dataset. Gaze error rate was modestly lower in adaptive
(18.2\%) than static (19.1\%) mode. The declutter mechanism reduced
timeouts (1.18\% → 0.42\%) but did not materially reduce slips (98.8\% →
99.6\%). Hand width inflation was not evaluable in the present dataset
(width scaling remained at 1.0; see Appendix). The declutter effect was
modest.

\subsection{Subjective Workload (RQ2)}\label{subjective-workload-rq2}

NASA-TLX scores (0--100) were higher for gaze than hand across all
subscales (Table~\ref{tbl-tlx}). Overall workload (unweighted mean of
six subscales) was 38.9 {[}35.3, 42.5{]} for hand and 46.4 {[}42.8,
50.0{]} for gaze. For the RQ2-specified subscales, \textbf{Physical
Demand} was 33.2 {[}29.4, 37.0{]} (hand) vs.~41.1 {[}37.1, 45.0{]}
(gaze), and \textbf{Frustration} was 31.4 {[}27.5, 35.4{]} (hand)
vs.~43.6 {[}39.6, 47.7{]} (gaze).

\begin{longtable}[]{@{}lcc@{}}
\caption{NASA-TLX subscale means {[}95\% CI{]} by modality. Higher =
more workload.}\label{tbl-tlx}\tabularnewline
\toprule\noalign{}
NASA-TLX Subscale & Hand & Gaze \\
\midrule\noalign{}
\endfirsthead
\toprule\noalign{}
NASA-TLX Subscale & Hand & Gaze \\
\midrule\noalign{}
\endhead
\bottomrule\noalign{}
\endlastfoot
Mental Demand & 33.8 {[}30.1, 37.4{]} & 45.1 {[}41.4, 48.9{]} \\
Physical Demand & 33.2 {[}29.4, 37.0{]} & 41.1 {[}37.1, 45.0{]} \\
Temporal Demand & 38.9 {[}35.3, 42.5{]} & 46.5 {[}43.1, 50.0{]} \\
Performance & 54.3 {[}48.6, 59.9{]} & 52.7 {[}48.4, 57.0{]} \\
Effort & 38.3 {[}34.4, 42.3{]} & 47.1 {[}43.3, 51.0{]} \\
Frustration & 31.4 {[}27.5, 35.4{]} & 43.6 {[}39.6, 47.7{]} \\
\textbf{Overall} & \textbf{38.9 {[}35.3, 42.5{]}} & \textbf{46.4
{[}42.8, 50.0{]}} \\
\end{longtable}

\begin{figure}[H]

\centering{

\includegraphics[width=0.7\linewidth,height=\textheight,keepaspectratio]{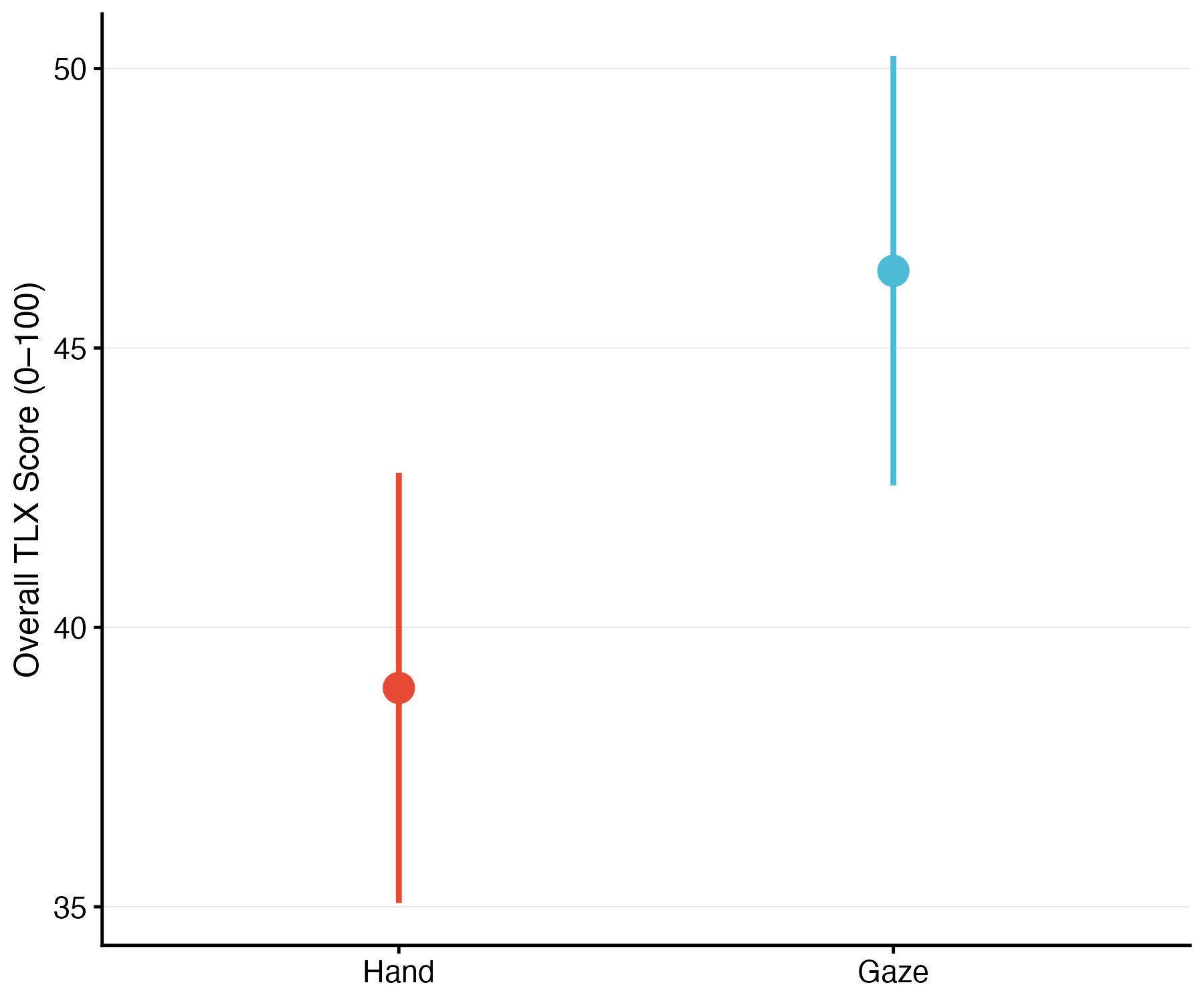}

}

\caption{\label{fig-tlx}NASA-TLX overall workload (0--100) by modality.
Gaze imposed higher subjective workload than hand. Error bars show 95\%
CI.}

\end{figure}

\begin{figure}[H]

\centering{

\includegraphics[width=0.75\linewidth,height=\textheight,keepaspectratio]{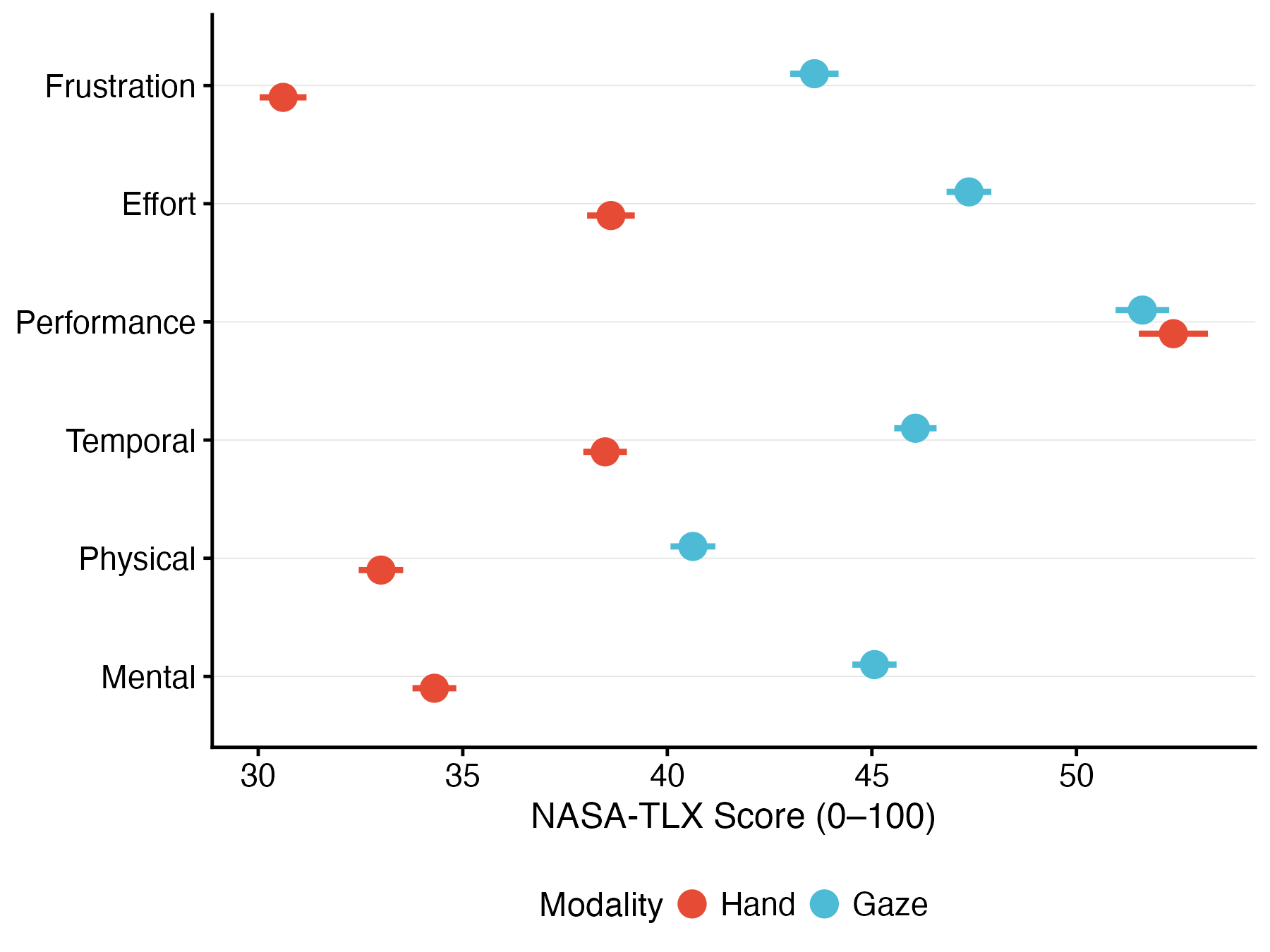}

}

\caption{\label{fig-tlx-subscales}NASA-TLX subscales by modality.
Point-range plot comparing Hand and Gaze on each subscale (Mental,
Physical, Temporal, Performance, Effort, Frustration). Gaze shows higher
scores on most dimensions; Physical Demand and Frustration show the
largest modality differences. Error bars show 95\% CI.}

\end{figure}

Figure~\ref{fig-tlx} shows overall workload by modality.
Figure~\ref{fig-tlx-subscales} compares workload across the six NASA-TLX
subscales, making it clear which dimensions (e.g., Physical Demand,
Frustration) differ most between hand and gaze.

\subsection{LBA Cognitive Modeling}\label{lba-cognitive-modeling}

\begin{figure}[H]

\centering{

\includegraphics[width=0.9\linewidth,height=\textheight,keepaspectratio]{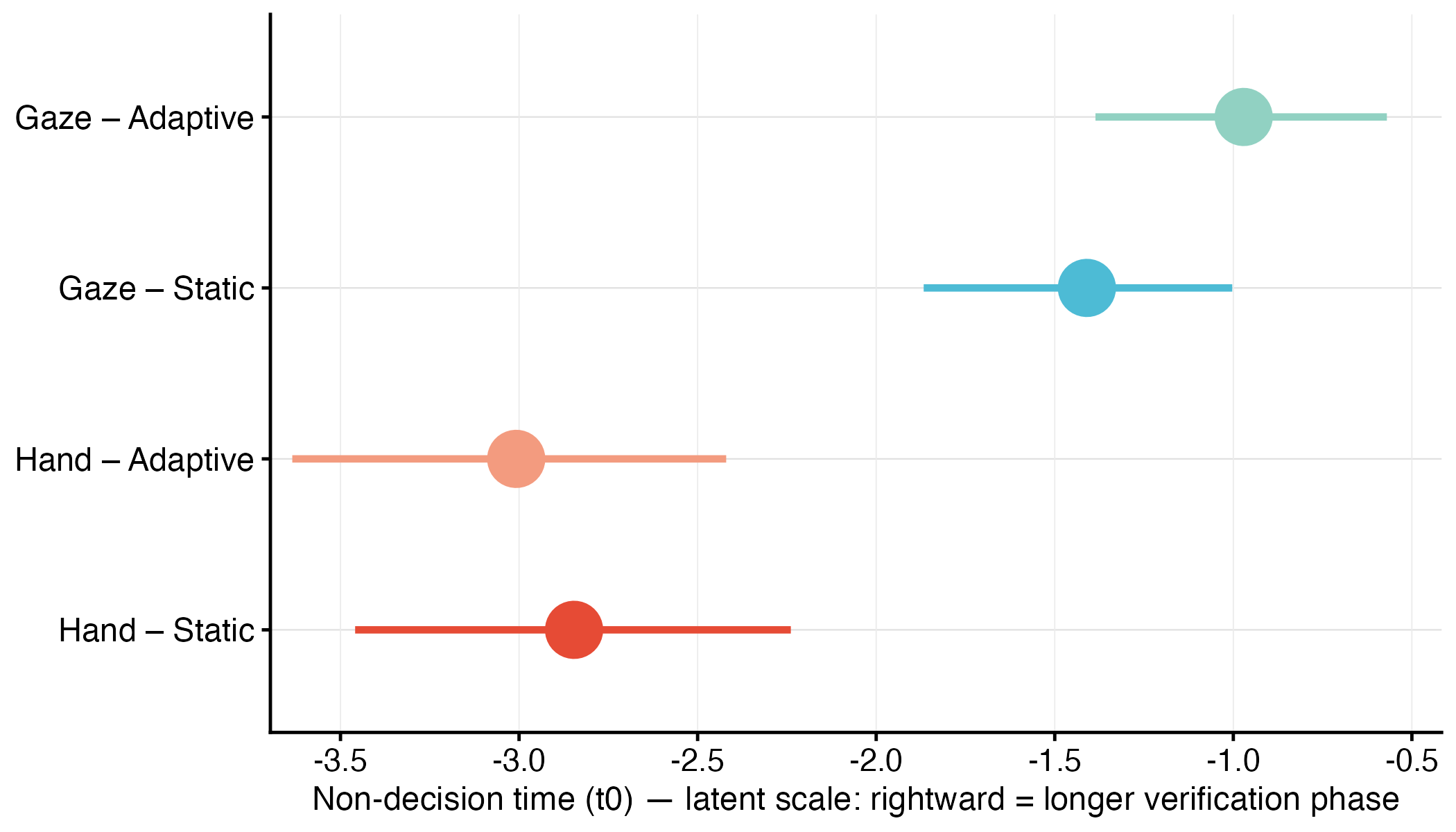}

}

\caption{\label{fig-lba-t0}Non-decision time (\(t_0\)) by condition.
Each row shows one condition; the point is the posterior mean and the
horizontal line is the 95\% highest-density interval. \(t_0\) is shown
on the model's latent scale, not in raw milliseconds. More rightward
(less negative) values correspond to a larger verification-related
offset under the fitted parameterization. Gaze conditions (top two rows)
show longer verification-related offsets than hand conditions (bottom
two rows), consistent with the empirical verification-phase RT and the
Midas Touch account.}

\end{figure}

\begin{figure}[H]

\centering{

\includegraphics[width=0.85\linewidth,height=\textheight,keepaspectratio]{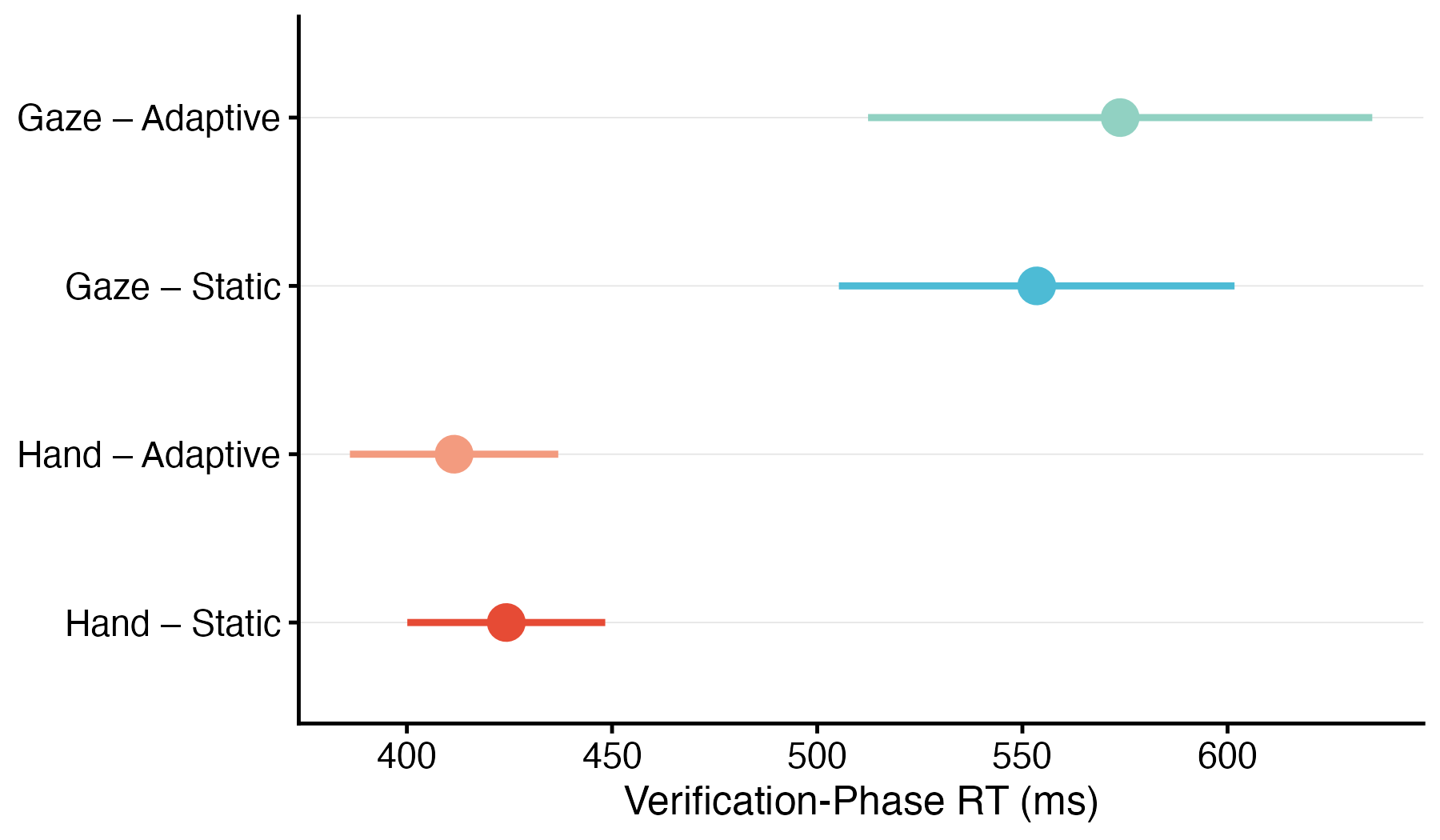}

}

\caption{\label{fig-verification-rt}Empirical verification-phase RT by
condition. Mean verification time (ms) from first target entry to
selection, with 95\% CI. The condition ordering in this figure matches
the latent \(t_0\) ordering in Figure~\ref{fig-lba-t0}, supporting the
interpretation that gaze conditions impose a longer verification-phase
burden than hand conditions.}

\end{figure}

Table~\ref{tbl-lba-params} reports group-level LBA parameter estimates
by modality and UI mode. In the fitted model, the primary
condition-sensitive parameter was \(t_0\); drift base, the ID-related
drift slope, and the pressure-related threshold slope were modeled as
shared effects and therefore do not vary across modality × UI rows in
Table~\ref{tbl-lba-params}. The non-decision-time parameter (\(t_0\)) is
reported on the model's latent scale, not as a raw duration in
milliseconds. More rightward (less negative) values indicate a larger
verification-related offset under the fitted parameterization.
Figure~\ref{fig-lba-t0} visualizes the \(t_0\) posterior estimates and
95\% HDIs by condition; gaze conditions show higher (less negative)
\(t_0\) values than hand conditions, consistent with longer
verification-phase duration in the empirical data.
Figure~\ref{fig-verification-rt} shows the empirical verification-phase
RT by condition; the condition ordering matches the latent \(t_0\)
ordering.

\subsubsection{Parameter Estimates}\label{parameter-estimates}

\begin{longtable}[]{@{}
  >{\raggedright\arraybackslash}p{(\linewidth - 8\tabcolsep) * \real{0.2000}}
  >{\centering\arraybackslash}p{(\linewidth - 8\tabcolsep) * \real{0.2200}}
  >{\centering\arraybackslash}p{(\linewidth - 8\tabcolsep) * \real{0.1200}}
  >{\centering\arraybackslash}p{(\linewidth - 8\tabcolsep) * \real{0.2200}}
  >{\centering\arraybackslash}p{(\linewidth - 8\tabcolsep) * \real{0.2400}}@{}}
\caption{LBA parameter estimates by modality and UI mode. Point
estimates with 95\% highest-density intervals. \(t_0\) (latent):
verification-related offset on the model's latent scale; values are not
in milliseconds. More rightward (less negative) values indicate longer
verification-phase duration under the fitted parameterization. ID slope:
effect of difficulty on drift (negative = harder trials reduce drift).
Pressure slope: effect of time pressure on threshold. Drift base and
threshold intercept shared across
conditions.}\label{tbl-lba-params}\tabularnewline
\toprule\noalign{}
\begin{minipage}[b]{\linewidth}\raggedright
Condition
\end{minipage} & \begin{minipage}[b]{\linewidth}\centering
\(t_0\) (latent) {[}95\% HDI{]}
\end{minipage} & \begin{minipage}[b]{\linewidth}\centering
Drift Base
\end{minipage} & \begin{minipage}[b]{\linewidth}\centering
ID Slope {[}95\% HDI{]}
\end{minipage} & \begin{minipage}[b]{\linewidth}\centering
Pressure {[}95\% HDI{]}
\end{minipage} \\
\midrule\noalign{}
\endfirsthead
\toprule\noalign{}
\begin{minipage}[b]{\linewidth}\raggedright
Condition
\end{minipage} & \begin{minipage}[b]{\linewidth}\centering
\(t_0\) (latent) {[}95\% HDI{]}
\end{minipage} & \begin{minipage}[b]{\linewidth}\centering
Drift Base
\end{minipage} & \begin{minipage}[b]{\linewidth}\centering
ID Slope {[}95\% HDI{]}
\end{minipage} & \begin{minipage}[b]{\linewidth}\centering
Pressure {[}95\% HDI{]}
\end{minipage} \\
\midrule\noalign{}
\endhead
\bottomrule\noalign{}
\endlastfoot
Hand -- Static & −2.85 {[}−3.46, −2.24{]} & 5.03 & −0.93 {[}−0.95,
−0.92{]} & 0.06 {[}0.03, 0.09{]} \\
Hand -- Adaptive & −3.01 {[}−3.64, −2.42{]} & 5.03 & −0.93 {[}−0.95,
−0.92{]} & 0.06 {[}0.03, 0.09{]} \\
Gaze -- Static & −1.41 {[}−1.87, −1.00{]} & 5.03 & −0.93 {[}−0.95,
−0.92{]} & 0.06 {[}0.03, 0.09{]} \\
Gaze -- Adaptive & −0.97 {[}−1.39, −0.57{]} & 5.03 & −0.93 {[}−0.95,
−0.92{]} & 0.06 {[}0.03, 0.09{]} \\
\end{longtable}

\subsubsection{Interpretation}\label{interpretation}

\textbf{Modality effect on \(t_0\):} Gaze conditions show higher \(t_0\)
values (less negative: −1.41 to −0.97) than hand conditions (−3.01 to
−2.85) on the latent scale, consistent with a longer verification-phase
duration for gaze interaction. This ordering aligns with the empirical
verification-phase RT (Figure~\ref{fig-verification-rt}). The pattern is
compatible with the Midas Touch account: gaze may require additional
time for intent disambiguation before selection.

\textbf{UI mode effect:} Adaptive UI shows different \(t_0\) patterns,
particularly for gaze (static: −1.41, adaptive: −0.97). The higher
\(t_0\) in gaze-adaptive may reflect altered verification timing under
declutter, though the direction warrants further investigation.

\textbf{Difficulty and pressure:} The shared effects capture general
task dynamics. The negative ID-related drift slope (−0.93) indicates
that greater task difficulty reduces evidence-accumulation efficiency
(consistent with Fitts's Law). The positive pressure-related threshold
slope (0.06) indicates that time pressure alters response caution---the
amount of evidence required before committing---rather than evidence
quality. The main modality difference remains concentrated in \(t_0\).

\subsubsection{Convergence Diagnostics}\label{convergence-diagnostics}

MCMC convergence was satisfactory: R-hat remained at 1.0 for all
parameters, and ESS exceeded 2,000. The drift-rate slope (ID) and
threshold slope (pressure) exhibited unimodal posteriors and well-mixed
chains. Non-decision time showed multiple modes across the four
condition cells, reflecting the hierarchical structure. Trace plots are
provided in the Appendix (Figure~\ref{fig-lba-trace}).

\section{Discussion}\label{discussion}

The results suggest a clear modality asymmetry in XR pointing
performance. Across the primary performance measures, hand input
outperformed gaze input: hand produced higher throughput (5.17 vs.~4.73
bits/s), lower error (1.8\% vs.~19.1\%), and shorter movement time (1.09
vs.~1.19 s). The workload pattern paralleled these performance
differences. NASA-TLX scores were higher for gaze than hand overall
(46.4 vs.~38.9), with especially notable differences in Physical Demand
(41.1 vs.~33.2) and Frustration (43.6 vs.~31.4). Taken together, these
findings address RQ1 and RQ2 in a consistent direction: under the
present task and simulation conditions, gaze interaction was not merely
less accurate than hand input, but also experienced as more effortful.

The most informative result is not only that gaze performed worse, but
how it failed. Gaze errors were overwhelmingly slips (99.2\%), whereas
hand errors were overwhelmingly misses (95.7\%). This asymmetry is
important because it distinguishes two different failure regimes. Hand
failures reflected spatial targeting difficulty: users attempted the
correct action but failed to acquire the target. Gaze failures instead
reflected intent ambiguity: the system registered activation when users
were looking, but not necessarily intending to select. This pattern is
directly consistent with the Midas Touch account proposed by Jacob
(1990). In that sense, the present dataset contributes more than another
hand-versus-gaze comparison. It shows that the principal liability of
gaze in this XR task was not generic imprecision alone, but the coupling
of visual attention and command execution.

These findings also fit the information-theoretic interpretation of
pointing performance formalized by Fitts's Law (Soukoreff \& MacKenzie,
2004). Throughput reflects the effective rate at which a user can
resolve spatial uncertainty, expressed in bits per second. On that
metric, hand preserved a higher information-processing rate than gaze in
the current task. The appendix-level Fitts validation supports this
interpretation: hand showed stronger fits and more stable scaling with
index of difficulty, whereas gaze exhibited weaker fits and lower
explained variance. Although both modalities were affected by task
difficulty, the steeper and noisier gaze fits suggest that its
performance was shaped by more than ballistic movement alone. In other
words, increasing difficulty did not simply elongate transport time; it
also appears to have amplified downstream verification demands.

The cognitive modeling results sharpen that interpretation. In the LBA
analysis, gaze showed a larger verification-related latent offset than
hand (\(t_0\) on the model's latent scale), indicating a longer pre- or
post-decisional component surrounding overt selection (Brown \&
Heathcote, 2008). This ordering was consistent with the empirical
verification-phase RT. In the present task, the most plausible
interpretation is that gaze required a longer verification phase before
commitment. That is exactly where a Midas Touch problem should appear:
not necessarily in the initial orientation toward a target, but in the
added delay needed to decide whether fixation is sufficiently stable,
intentional, and safe to confirm. The negative ID slope for drift rate
(−0.93) is also consistent with a Fitts-like account in which harder
trials reduce the rate of evidence accumulation. The positive pressure
slope (0.06) further suggests that time pressure altered the caution
policy rather than simply speeding responses, which is consistent with a
speed--accuracy tradeoff account. Taken together, the LBA results
indicate that gaze interaction imposed a verification burden beyond raw
target acquisition. Because the condition-varying effect was
concentrated in \(t_0\), the modality difference in this dataset appears
to reflect a verification burden more than a broad change in evidence
quality or response caution.

This pattern can also be understood through the lens of cognitive load.
In principle, gaze offers a low-effort means of orienting to targets,
while hand input incurs greater bodily effort, especially in spatial
interfaces often associated with ``Gorilla Arm'' fatigue. Yet the
present workload results suggest that, in this task, gaze introduced
greater extraneous load than hand. That load likely came from the need
to monitor cursor stability, manage dwell or confirmation timing, and
suppress unintended selections. Hand input, by contrast, appears to have
shifted the burden toward controlled spatial targeting, but in a way
that remained more manageable under the current task constraints. In
that sense, the gaze--hand tradeoff is not well captured by a simple
speed-versus-fatigue dichotomy. Rather, gaze may reduce some motor
demands while increasing decisional and attentional overhead; hand may
require more overt motor control while preserving clearer action
intention and lower ambiguity. The workload findings are consistent with
that framing, particularly the elevated Frustration scores under gaze.

RQ3 received only partial support in this dataset. The only adaptive
mechanism that actually executed was gaze declutter, and its benefit was
modest. Gaze error declined directionally from 19.1\% in the static
condition to 18.2\% in the adaptive condition. More specifically,
declutter reduced timeouts from 1.18\% to 0.42\%, but did not reduce
slips; slips remained dominant and even slightly increased
proportionally (98.8\% to 99.6\%). This pattern suggests that declutter
may have helped with distraction-driven hesitation or delayed
commitment, but it did not address the core failure mode of gaze
interaction, namely intent disambiguation. That distinction matters: if
the main problem is that users accidentally activate while looking,
reducing peripheral clutter may improve attentional focus without
solving the actual selection ambiguity. By contrast, the hand
width-inflation mechanism was not evaluable in the present dataset
(width scaling remained at 1.0; the policy engine emitted actions in
some sessions but the UI did not apply them to rendered targets). RQ3 is
only partially answered given that hand width inflation was not
evaluable.

The present findings are broadly aligned with prior work showing both
the promise and fragility of gaze-based interaction in multimodal
systems. Prior gaze+hand paradigms have often treated gaze as an
efficient means of target acquisition and the hand as a confirmation or
refinement channel, thereby exploiting the complementary strengths of
the two modalities (Pfeuffer et al., 2017). That general logic is
compatible with the current results: gaze appears efficient for
orienting attention, but unreliable as a sole selection mechanism when
intent must be inferred from fixation. Likewise, the width-inflation
mechanism draws conceptually on expanding-target and Bubble Cursor work,
which has repeatedly shown that increasing effective target size can
improve pointing performance under uncertainty (Grossman \&
Balakrishnan, 2005; McGuffin \& Balakrishnan, 2005). A key contribution
of the present platform is the integration of physiologically
constrained gaze simulation, an ISO 9241-9 task structure, and
policy-driven modality-specific adaptive interventions within one
reproducible framework. That combination makes it possible to study
modality-specific failure modes under controlled conditions rather than
treating gaze and hand as interchangeable input channels.

The design implications are practical. For XR designers, the current
evidence suggests that hand input remains the more reliable option for
precise selection tasks in which slips are costly. Gaze may still be
valuable for rapid orienting, scanning, or coarse target acquisition,
but only when paired with a mechanism that resolves intent ambiguity.
Declutter may help when performance degradation is driven by visual
competition or hesitation, but it is unlikely to solve Midas Touch on
its own. More generally, adaptive systems in XR should not be judged
only by whether adaptation occurred, but by whether the adaptation
targets the dominant failure mode of the modality.
Table~\ref{tbl-design-implications} summarizes these modality-specific
guidelines.

\begin{longtable}[]{@{}
  >{\raggedright\arraybackslash}p{(\linewidth - 4\tabcolsep) * \real{0.1724}}
  >{\raggedright\arraybackslash}p{(\linewidth - 4\tabcolsep) * \real{0.3966}}
  >{\raggedright\arraybackslash}p{(\linewidth - 4\tabcolsep) * \real{0.4310}}@{}}
\caption{Design implications: modality-specific failure modes and
recommended adaptive
interventions.}\label{tbl-design-implications}\tabularnewline
\toprule\noalign{}
\begin{minipage}[b]{\linewidth}\raggedright
Modality
\end{minipage} & \begin{minipage}[b]{\linewidth}\raggedright
Dominant Failure Mode
\end{minipage} & \begin{minipage}[b]{\linewidth}\raggedright
Recommended Adaptation
\end{minipage} \\
\midrule\noalign{}
\endfirsthead
\toprule\noalign{}
\begin{minipage}[b]{\linewidth}\raggedright
Modality
\end{minipage} & \begin{minipage}[b]{\linewidth}\raggedright
Dominant Failure Mode
\end{minipage} & \begin{minipage}[b]{\linewidth}\raggedright
Recommended Adaptation
\end{minipage} \\
\midrule\noalign{}
\endhead
\bottomrule\noalign{}
\endlastfoot
Gaze & Slips (unintended activation) & Declutter for distraction-driven
hesitation; intent disambiguation (dwell, confirmation) for core Midas
Touch \\
Hand & Misses (spatial targeting) & Width inflation / target expansion
for acquisition difficulty \\
\end{longtable}

Several limitations qualify these conclusions. First, the sample
comprised N=69 participants with complete factorial data. Second, gaze
was implemented via a psychophysically-grounded simulation rather than
hardware eye-tracking (see Figure~\ref{fig-psychophysics}). This is a
deliberate methodological strength: parametric control over saccadic
suppression, fixation jitter, and sensor lag---each calibrated to
published oculomotor norms---provides reproducibility and internal
validity that hardware trackers cannot offer, since real systems
confound device-specific artifacts (calibration drift, illumination
variability, blink events) with genuine oculomotor signals. The
constraint of the simulation is scope, not accuracy: it models the three
psychophysical mechanisms most relevant to dwell-based selection, but
does not replicate the full noise profile of a specific eye-tracker
model. Whether the modality asymmetries documented here---particularly
the slip-dominated gaze error profile---persist under these additional
noise sources is an empirical question for follow-up work with hardware
eye-tracking. Third, the hand adaptation pathway was not evaluable
(width inflation did not affect rendered targets), leaving RQ3 only
partially answered. Fourth, a pressure-logging bug affected early data
collection; the primary exclusion criterion is 8-block factorial
completeness. Fifth, the study used a desktop web proxy
(mouse/trackpad); generalization to headset-based XR with optical hand
tracking or hardware eye-tracking requires validation. Finally, the task
was restricted to an ISO 9241-9 multidirectional tapping paradigm, which
captures one well-defined class of XR interaction but cannot fully
represent more ecological tasks such as menu navigation, object
manipulation, or extended mixed-modality workflows.

These limitations point to directions for future work. Validation with
real eye-tracking hardware could determine whether the present
gaze-specific costs persist under actual ocular input. Expansion of the
adaptive design space---for example, dynamic dwell policies, goal-aware
snapping, or hybrid confirmation schemes---could target intent
disambiguation more directly. The hand pathway should be evaluated under
conditions where width inflation activates reliably. Broader task
ecologies are needed: modality-specific adaptive systems will be most
compelling if they generalize beyond canonical tapping tasks to the
mixed attentional and motor demands that define real XR work.

\section{Conclusion}\label{conclusion}

This paper introduced the xr-adaptive-modality-2025 platform as a
rigorous and reproducible framework for studying gaze and hand pointing
and modality-specific adaptive interventions in XR-relevant tasks. The
platform combines an ISO 9241-9 multidirectional tapping task,
physiologically informed gaze simulation, and policy-driven adaptive
interventions intended to respond to modality-specific performance
degradation. As a research framework, it is designed not only to compare
gaze and hand input, but to examine when and why adaptation helps.

The findings from the factorial dataset (N=69) suggest a consistent
pattern. Hand input yielded higher throughput and lower error than gaze
input, and gaze imposed higher subjective workload across NASA-TLX
dimensions. The strongest empirical result was the error-type asymmetry:
gaze errors were almost entirely slips, whereas hand errors were
overwhelmingly misses. That pattern is consistent with the Midas Touch
problem (Jacob, 1990) and indicates that gaze failure in this task was
driven primarily by intent ambiguity rather than by timeout-based
hesitation alone. The only adaptive mechanism that executed---gaze
declutter---showed modest directional benefit by reducing timeouts, but
did not reduce slips. The hand width-inflation mechanism was not
evaluable in this dataset.

For XR designers, the practical implication is straightforward: adaptive
systems should be built around modality-specific failure modes rather
than generic adaptation logic. Declutter may help when gaze performance
degrades due to distraction, while target expansion may be more
appropriate for hand-based spatial difficulty, but both require
evaluation under the conditions that actually trigger them. Adaptive
multimodal XR interfaces will likely be most effective when they treat
gaze and hand as distinct channels with distinct ergonomic and cognitive
constraints.

\section{Code and Materials
Availability}\label{code-and-materials-availability}

Code, analysis scripts, and documentation are available at
https://github.com/mohdasti/xr-adaptive-modality-2025. The repository
includes the experimental platform (React/TypeScript), R and Python
analysis pipelines, preregistration documents, and data dictionaries.
Aggregated results are available via Zenodo (DOI:
10.5281/zenodo.18204915). The live deployment used for data collection
is hosted at https://xr-adaptive-modality-2025.vercel.app.

\section{Acknowledgments}\label{acknowledgments}

This research was conducted as an independent project. The authors thank
the participants for their time and the open-source community for the
tools that made this work possible.

\section{Appendix}\label{appendix}

\subsection{Fitts' Law Regression
(Validation)}\label{fitts-law-regression-validation}

Linear regression of movement time on effective Index of Difficulty
(\(ID_e\)) validates that the task conforms to Fitts's Law.
Table~\ref{tbl-fitts} reports slope (\(b\)) and \(R^2\) by modality and
UI mode. Hand conditions showed smaller slopes (0.15--0.16 s/bit) and
higher \(R^2\) (0.54) than gaze (slopes 0.18--0.19 s/bit, \(R^2\)
0.25--0.35). The steeper gaze slope suggests that difficulty primarily
affects the verification phase rather than the initial ballistic phase,
aligning with the LBA latent-offset results.

\begin{longtable}[]{@{}lcc@{}}
\caption{Fitts' Law regression (MT \textasciitilde{} \(ID_e\)) by
condition. Slope = movement-time cost per bit (smaller = higher
efficiency); \(R^2\) = proportion of variance
explained.}\label{tbl-fitts}\tabularnewline
\toprule\noalign{}
Condition & Slope (s/bit) & \(R^2\) \\
\midrule\noalign{}
\endfirsthead
\toprule\noalign{}
Condition & Slope (s/bit) & \(R^2\) \\
\midrule\noalign{}
\endhead
\bottomrule\noalign{}
\endlastfoot
Hand -- Static & 0.155 & 0.54 \\
Hand -- Adaptive & 0.146 & 0.54 \\
Gaze -- Static & 0.179 & 0.35 \\
Gaze -- Adaptive & 0.19 & 0.25 \\
\end{longtable}

\begin{figure}[H]

\centering{

\includegraphics[width=0.95\linewidth,height=\textheight,keepaspectratio]{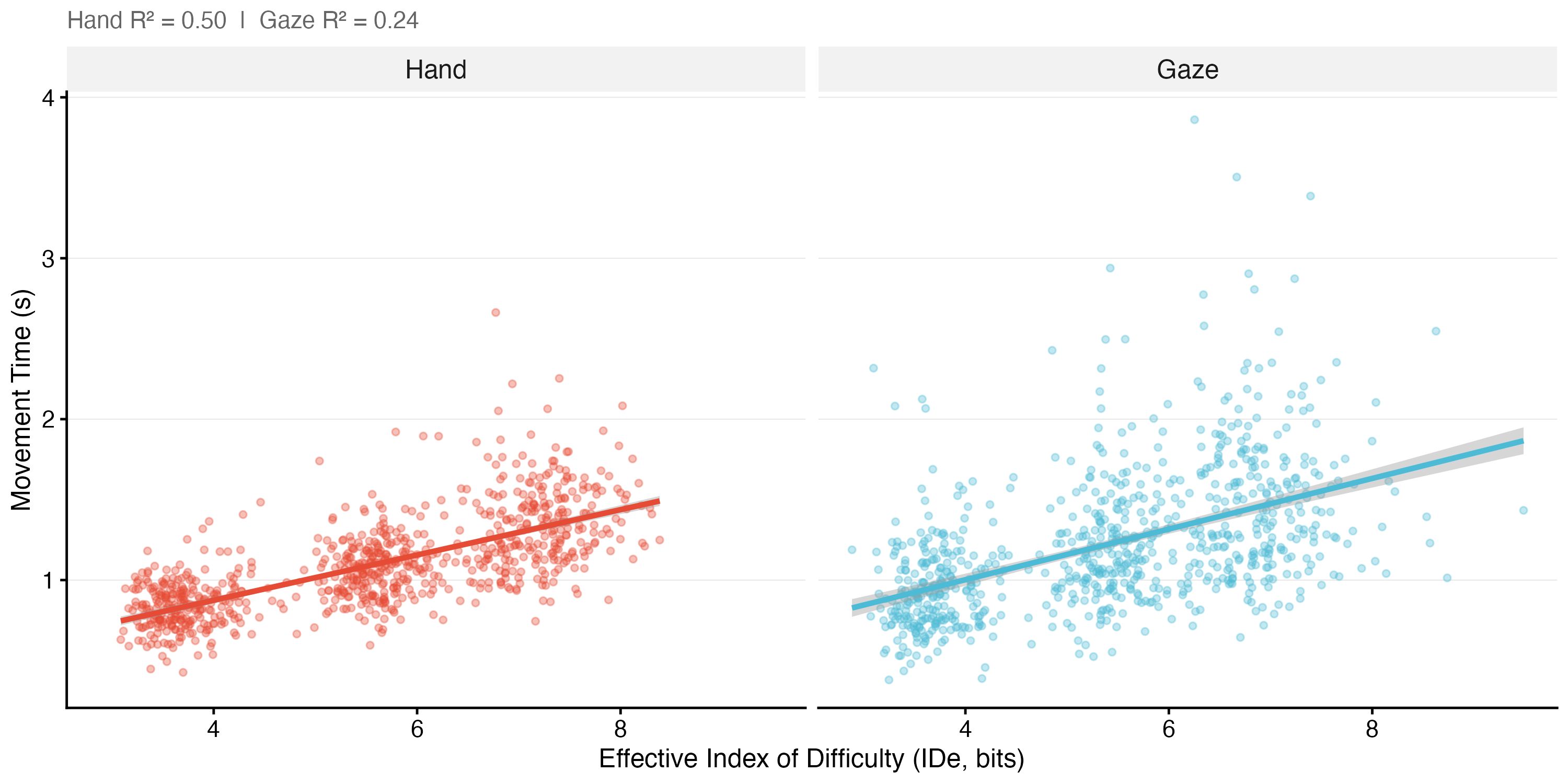}

}

\caption{\label{fig-fitts}Fitts' Law validation. Movement time
vs.~effective Index of Difficulty by modality. Linear fits with 95\% CI
bands.}

\end{figure}

Figure~\ref{fig-fitts} shows the regression of movement time on
effective Index of Difficulty by modality.

\subsection{LBA MCMC Trace Plots}\label{lba-mcmc-trace-plots}

\begin{figure}[H]

\centering{

\includegraphics[width=0.9\linewidth,height=\textheight,keepaspectratio]{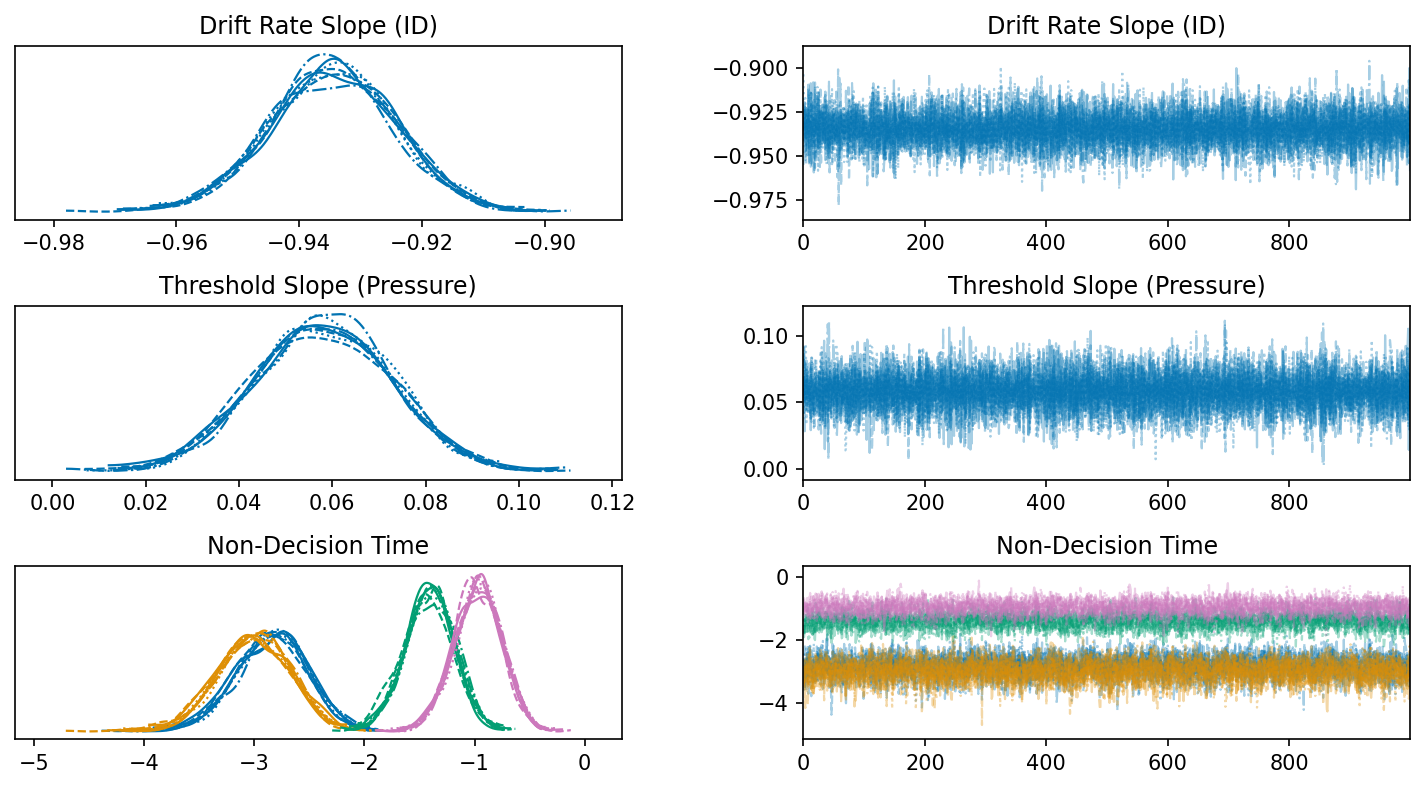}

}

\caption{\label{fig-lba-trace}MCMC trace plots for key LBA parameters.
Left panels: posterior densities; right panels: sampling chains. Drift
rate slope (ID) and threshold slope (pressure) show unimodal posteriors
and good mixing; non-decision time reflects the four condition-specific
estimates.}

\end{figure}%

\subsection{Adaptive System Manipulation
Check}\label{adaptive-system-manipulation-check}

Hand width inflation did not activate in this dataset. Across all
trials, \texttt{width\_scale\_factor} remained 1.0 (0 trials with
scaling). The PolicyEngine emitted \texttt{inflate\_width} actions in
some sessions, but the UI integration did not apply them to rendered
targets. Consequently, hand UI-mode effects cannot be interpreted as
adaptation effectiveness; only gaze declutter was evaluable. Full
diagnostic analysis is available in the project repository.

\section{References}\label{references}

\phantomsection\label{refs}

\end{document}